# Revealing spontaneous symmetry breaking in continuous time crystals

Yuanjiang Tang[1]†, Chenyang Wang[1]†, Bei Liu[2]†, Jin Peng[2], Chao Liang[1], Yaohua Li[1], Xian Zhao[2], Cuicui Lu[3,4], Shuang Zhang[4,5,6]*, Yong-Chun Liu[1,7]*

[1]State Key Laboratory of Low-Dimensional Quantum Physics, Department of Physics, Tsinghua University, Beijing 100084, China

[2]Key Laboratory of Laser & Infrared System, Ministry of Education, Center for Optics Research and Engineering (CORE), Shandong University, Qingdao 266237, China

[3]Key Laboratory of Advanced Optoelectronic Quantum Architecture and Measurements of Ministry of Education, Beijing Key Laboratory of Nanophotonics and Ultrafine Optoelectronic Systems, Center for Interdisciplinary Science of Optical Quantum and NEMS Integration, School of Physics, Beijing Institute of Technology, Beijing 100081, China

[4]Department of Physics, University of Hong Kong, Hong Kong 999077, China

[5]Department of Electrical and Electronic Engineering, University of Hong Kong, Hong Kong, 999077, China

[6]Quantum Science Center of Guangdong-Hong Kong-Macao Great Bay Area, 3 Binlang Road, Shenzhen, China

[7]Frontier Science Center for Quantum Information, Beijing 100084, China

*Corresponding author. Email: shuzhang@hku.hk (S. Zhang); ycliu@tsinghua.edu.cn (Y.-C. Liu)

†These authors contributed equally to this work.

**Abstract:** Spontaneous symmetry breaking plays a pivotal role in physics ranging from the emergence of elementary particles to the phase transitions of matter. The spontaneous breaking of continuous time translation symmetry leads to a novel state of matter named continuous time crystal (CTC). It exhibits periodic oscillation without the need for periodic driving, and the relative phases for repetitively realized oscillations are random. However, the mechanism behind the spontaneous symmetry breaking in CTCs, particularly the random phases, remains elusive. Here we propose and experimentally realize two types of CTCs based on distinct mechanisms: manifold topology and near-chaotic motion. We observe both types of CTCs in thermal atomic ensembles by artificially synthesizing spin-spin nonlinear interactions through a measurement-feedback scheme. Our work provides general recipes for the realization of CTCs, and paves the way for exploring CTCs in various systems.

Spontaneous symmetry breaking, which refers to the phenomenon that a system's state does not exhibit the same symmetry as the underlying laws that govern it, is one of the cornerstones of modern physics. It is the core of the Higgs mechanism of the mass origin for elementary particle, the Landau theory of phase transitions, and the formation of crystals. Recently, the study of spontaneous symmetry breaking in the temporal dimension, which results in the concept of time crystal, has opened new frontiers for exploring physical laws beyond the spatial dimensions. The concept was initially introduced by Shapere and Wilczek to describe the lowest energy state of a closed system exhibiting periodic oscillations under the action of time-invariant Hamiltonian (*1–4*). However, it was later noted that the no-go theorem prohibits the existence of time crystals in closed equilibrium systems(*5, 6, 7*). To circumvent the restrictions, attentions were paid to the non-equilibrium and open systems(*8–14*). Discrete time crystals, which break the discrete time translation symmetry, have been discovered in periodically-driven (Floquet) non-equilibrium systems (*15–37*). In open systems, continuous time crystals (CTCs), which spontaneous breaks the continuous time translation symmetry, have been achieved under many-body or nonlinear interactions(*38–44*).

In CTC, the physical laws that govern the system, like Hamiltonian or equations of motion, does not change with time, satisfying the continuous time translation symmetry. However, the state of the system spontaneous breaks the continuous time translation symmetry, exhibiting stable periodic oscillation with random initial phase (Fig. 1A). A prominent example of the periodic oscillation is the limit cycle, which refers to an isolated period orbit in phase space. To ensure that the symmetry breaking process occurs spontaneously, the oscillations should exist at arbitrary starting time, corresponding to random initial phases for repetitively realized oscillations (*39, 45*). However, the mechanisms for generating the random phase, which is the most crucial condition for the CTCs, remain unknown. This represents significant challenges for the realization of CTCs.

In this work, we reveal the mechanisms of spontaneous breaking of continuous time translation symmetry. We propose two schemes to achieve CTCs by generating random phases within the limit cycle phase. For the type-I CTC, the random phases are ensured by the connected topology of the unstable manifold of the unstable fixed points. For the type-II CTC, the random phases arise from the near-chaotic hopping between unstable periodic orbits, which amplify the initial fluctuations. We experimentally observe these two types of CTCs in a thermal atomic ensemble with artificial spin-spin interactions by using the measurement-feedback method. This work opens new ways to realize CTCs, and brings more opportunities for exploring CTCs in various systems.

The proposed type-I CTC is ensured by the existence of two or higher dimensional unstable manifold. As schematically illustrated in Fig. 1B, we take an example of a system with a two-dimensional unstable manifold $W_u$ (yellowish surface) and a one-dimensional stable manifold $W_s$ (arrowed solid line). Near the unstable fixed point, the flows will be attracted to $W_u$ along the stable manifold $W_s$, and move repulsively along the unstable manifold $W_u$, as shown by the blue curves in Fig. 1B. When the state of the system is initialized to the unstable fixed point, fluctuations will lead to significant difference in the trajectories, resulting in the randomness of the phases. Because of the attractive dynamics towards $W_u$, only the fluctuations tangent to $W_u$ are effective in the random phase generation. When the dimensionality of the unstable manifold is two or higher, the fluctuations tangent to the unstable manifolds are continuously distributed in all directions, which generates the random phases ranging continuously from 0 to $2\pi$ after the trajectories reach the LC. As a contrast, when the unstable manifold is one-dimensional, the flows moves along the two directions of the unstable manifold to the LC, and the phases show a two-point distribution (see the SM).

The proposed Type-II CTCs appear when the near-chaotic motion occurs between multiple unstable periodic orbits (Fig. 1C). The trajectories of states (gradual purple curves) are firstly attracted to the neighborhood of one unstable periodic orbit (red orbit), then move repulsively along the unstable manifold $W_u$. After leaving the neighborhood of the periodic orbit, the trajectories randomly jump among the unstable periodic orbits, until they are attracted to the LC and keep stable periodic motion. When the repulsion of the unstable periodic orbits is weak, the flows near the unstable periodic orbits exhibit near-stable motion, which elongates the time duration of the near-chaotic motion. As a result, tiny initial fluctuations will cause significant time differences when the trajectories enter the LC, resulting in the $2\pi$ random phase distribution.

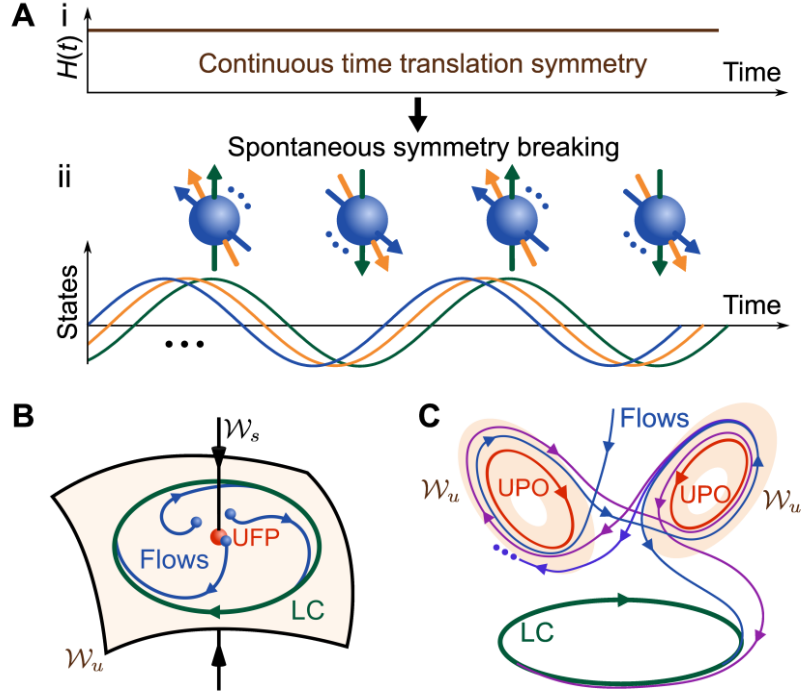

**Fig.1. Overview of the characteristics and the proposed two realizations of CTCs.** (**A**) The characteristics of the CTC phase, where (i) shows the time-invariant Hamiltonian and (ii) shows the oscillating states with random phases. The solid lines of different colors indicate that the initial phases of the oscillations are random. (**B**) The schematic diagram of the type-I CTC. The red dot marked by 'UFP' denotes the unstable fixed point. The stable manifold $W_s$ and unstable manifold $W_u$ are denoted by the arrowed solid line and the yellowish surface, respectively. The blue solid dots are the initial values where the noise perturbations make small differences, and they can evolve to the limit cycle in any direction along the two-dimensional surface, resulting in a random phase from 0 to $2\pi$. (**C**) The schematic diagram of the type-II CTC. The red and green solid loops denote the unstable periodic orbits (marked by 'UPO') and the LC, respectively. The flows with slight differences in initial values jump near-chaotically between the unstable periodic orbits, and the small initial value differences in the flows lead to the different evolution time in the unstable periodic orbits, as shown by the four gradual purple solid lines. Finally, all flows are attracted to the limit cycle, but the phase is distributed randomly from 0 to $2\pi$.

We observe both of the two types of CTCs in a rubidium atomic ensemble, with the experimental setup shown in Fig.2A. A cube glass cell is filled with an ensemble of thermal rubidium atoms, and the atomic states can be described by a collective spin with spin polarization $\mathbf{P} = (P_x, P_y, P_z)$, where $P_{\mu=x,y,z}$ is the spin polarization component along $\mu$ axis. A beam of circularly polarized laser propagating along the $z$ direction optically pumps the atomic ensemble to polarize the collective spin. We measure the spin polarization component $P_x$ using a probe laser via balanced detection. The output signal is fed into a loop that includes a feedback resistor and the $y$-axis feedback coil, which generates the feedback magnetic field $B_y = -\Gamma_{FB} P_x / \gamma$ with $\Gamma_{FB}$ being the feedback factor and $\gamma$ being the gyromagnetic ratio of atoms (*46*, *47*). A static magnetic field of $B_0$ is applied along the $z$ axis, and the dynamical evolution of the spin polarization can be described by the Bloch equations,

$$\begin{cases} \dfrac{d}{dt} P_x = -\dfrac{P_x}{T_2} + \gamma B_0 P_y + \Gamma_{FB} P_x P_z \\ \dfrac{d}{dt} P_y = -\gamma B_0 P_x - \dfrac{P_y}{T_2} \\ \dfrac{d}{dt} P_z = -\Gamma_{FB} P_x^2 - \dfrac{P_z}{T_1} + R_{op} \end{cases}, \qquad (1)$$

where $T_1$ ($T_2$) is the longitudinal (transverse) relaxation time, and $R_{op}$ is the optical pumping rate. Note that the feedback magnetic field $B_y$ carries the information of the spin polarization component $P_x$, which will result in equivalent spin-spin interaction terms $P_x P_z$ and $P_x^2$. By adjusting the feedback factor, we can modulate the strength of the nonlinear spin-spin interactions. Different combinations of feedback factors and static magnetic field $B_0$ produce various nonlinear dynamic phases.

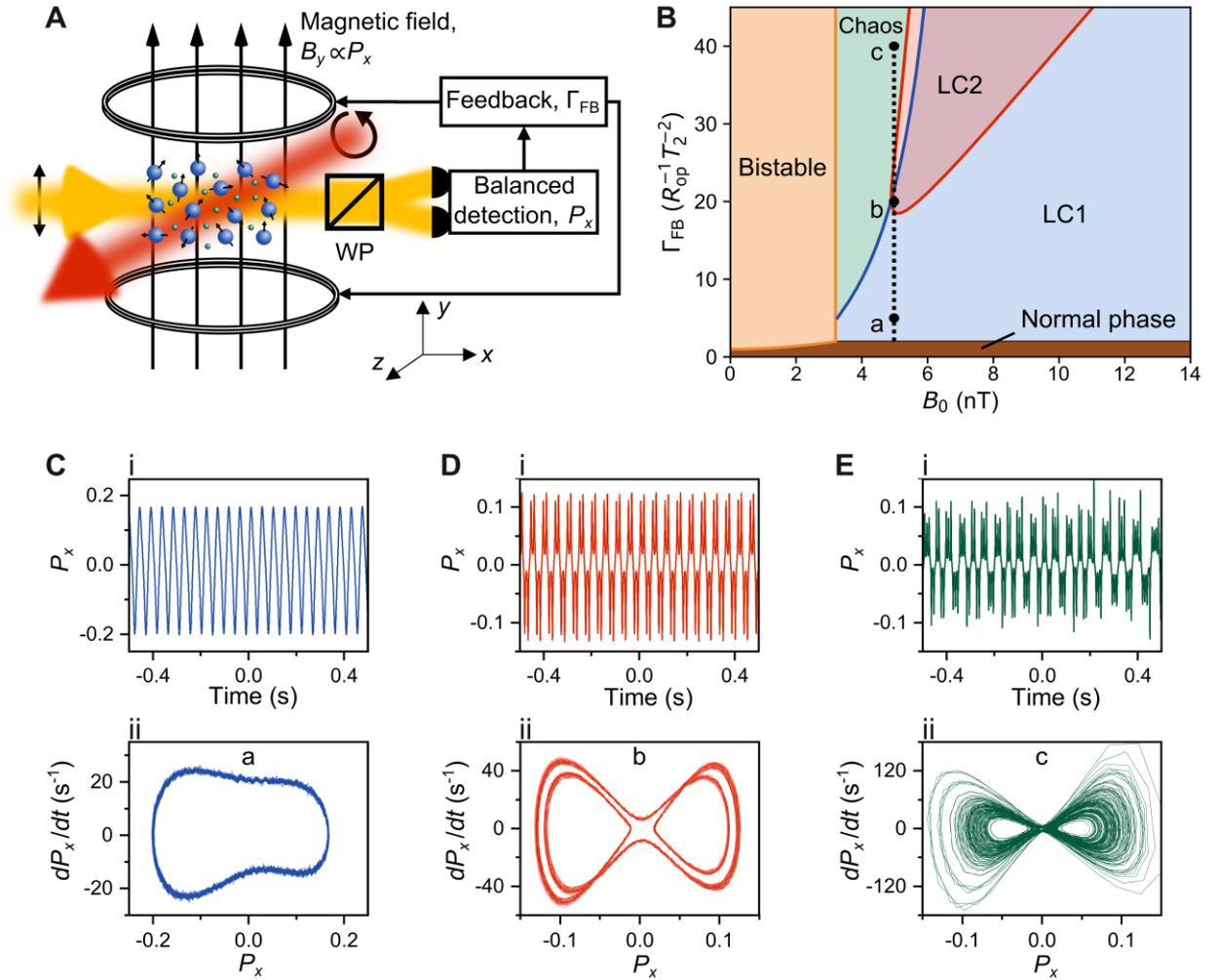

**Fig.2. Experimental setup, phase diagrams and representative nonlinear dynamics.**
**(A)** The experimental setup. A circularly polarized pump laser (red arrow) with power being 5 mW and frequency being 10 GHz above the $^{87}$Rb D1 line is illuminated to a thermal $^{87}$Rb atomic ensemble along $z$-axis direction to polarize the atoms. A linearly polarized probe laser propagating along $x$-axis direction (yellow arrow) with power being 50 $\mu$W and frequency being 230 GHz above the D1 line is used to detect the $P_x$ component of the spin polarization. The detected signal is transformed into the magnetic field along $y$-axis direction through a feedback loop. **(B)** Phase diagram of the nonlinear dynamics for different magnetic fields $B_0$ and feedback factors $\Gamma_{FB}$. The system has a normal phase (brown region), bistable phase (yellow region), limit cycle phases (blue and red regions), and a chaotic phase (green region). LC1 and LC2 represent different types of limit cycles. Points a, b and c are located in the LC1, LC2 and chaotic phases, respectively. **(C)** Output signals (i) and phase space trajectories of

$P_x$ (ii) at $B_0 = 4.98$ nT, $\Gamma_{FB} = 5R_{op}^{-1}T_2^{-2}$ (point a). The system is in the LC1 region.

**(D)** Point b, $B_0 = 4.98$ nT, $\Gamma_{FB} = 20R_{op}^{-1}T_2^{-2}$. The system is in the LC2 region. **(E)** Point c, $B_0 = 4.98$ nT, $\Gamma_{FB} = 40R_{op}^{-1}T_2^{-2}$. The system is in the chaos region.

The nonlinear dynamics of the spin polarization vary significantly under different experimental parameters, including the formation of fixed points, limit cycles and chaotic dynamics. We theoretically and numerically investigate the nonlinear dynamical phases, and experimentally observe the presence of different phases, as shown in Fig. 2. Fig. 2B shows the phase diagram of the nonlinear dynamics under varying static field $B_0$ and strength of feedback $\Gamma_{FB}$. Under weak feedback, the system exhibits a normal phase (brown region), where the collective spin is polarized to a steady state $\mathbf{P}^{(0)} = (0, 0, R_{op}T_1)$ as in Eq. 1, resembling the linear dynamics without feedback. By increasing the feedback strength, the state $\mathbf{P}^{(0)}$ becomes unstable, and novel dynamical properties can emerge. When $B_0 < 1/\gamma T_2$, the steady state $\mathbf{P}^{(0)}$ bifurcates into two stable fixed points as feedback strength increases, and the system enters a bistable phase (see the SM).

When $B_0 > 1/\gamma T_2$, a Hopf bifurcation occurs with an increase in the feedback strength, and $\mathbf{P}^{(0)}$ bifurcates into a limit cycle. Under limit cycle phases, the spin polarization oscillates periodically in long-time dynamics, and the trajectory in the phase space remains unchanged under small perturbations of the initial polarizations. By sweeping the parameters, we find that different types of limit cycles appear and disappear under different parameters, and limit cycles may coexist, depending on the initial polarizations. As shown in Fig. 2B, two different limit cycles are found in our experimental parametric range, labeled as the limit cycle 1 (LC1, blue region) and limit cycle 2 (LC2, red region). The overlap of the regions for different limit cycles in Fig. 2B means the coexistence of different limit cycles. When all the limit cycles are unstable, shown as the green region in Fig. 2B, the system shows chaotic dynamics, that is, the long-time dynamics show random jumps among different unstable periodic orbits (see the SM).

To illustrate the properties of different limit cycles and chaos, we select points 'a' ($\Gamma_{FB} = 5R_{op}^{-1}T_2^{-2}$), 'b' ($\Gamma_{FB} = 20R_{op}^{-1}T_2^{-2}$), and 'c' ($\Gamma_{FB} = 40R_{op}^{-1}T_2^{-2}$) in LC1, LC2, and the chaotic phase, respectively, under a fixed $B_0 = 4.98$ nT. The probe signal of $P_x$ oscillating with time is measured, and the phase space trajectory consisting of $P_x$ and its derivative $dP_x/dt$ is calculated as shown in Fig. 2C-E. Point 'a' is in the LC1 region, and signal $P_x$ oscillates periodically with one peak per half cycle (Fig. 2C(i)). The corresponding phase space trajectory is similar to the contour of a butterfly's wing as in Fig. 2C(ii). Point 'b' has a larger feedback coefficient than point 'a'. At point 'b', the system will end up in LC2 when starting from the initial spin polarization before the feedback turned on. As shown in Fig. 2D(i), LC2 exhibits two peaks per half cycle. This results in two more peaks in LC2 than LC1 in the complete cycle, and thus, two ring structures appear within the butterfly contour of the phase space trajectory, as shown in Fig. 2D(ii). The feedback factor continues to increase, the system enters the chaotic phase where point 'c' is located. In the chaotic phase, all the periodic orbits lose their stability, and the system jumps between different periodic orbits, producing intermittent chaos. As shown in Fig. 2E(i), the variation of $P_x$ is non-periodic, producing the random appearance of double, triple, or even more peaks. Furthermore, the phase space trajectories are non-closed and feature a complex structure of alternating multiple rings, as shown in Fig. 2E(ii).

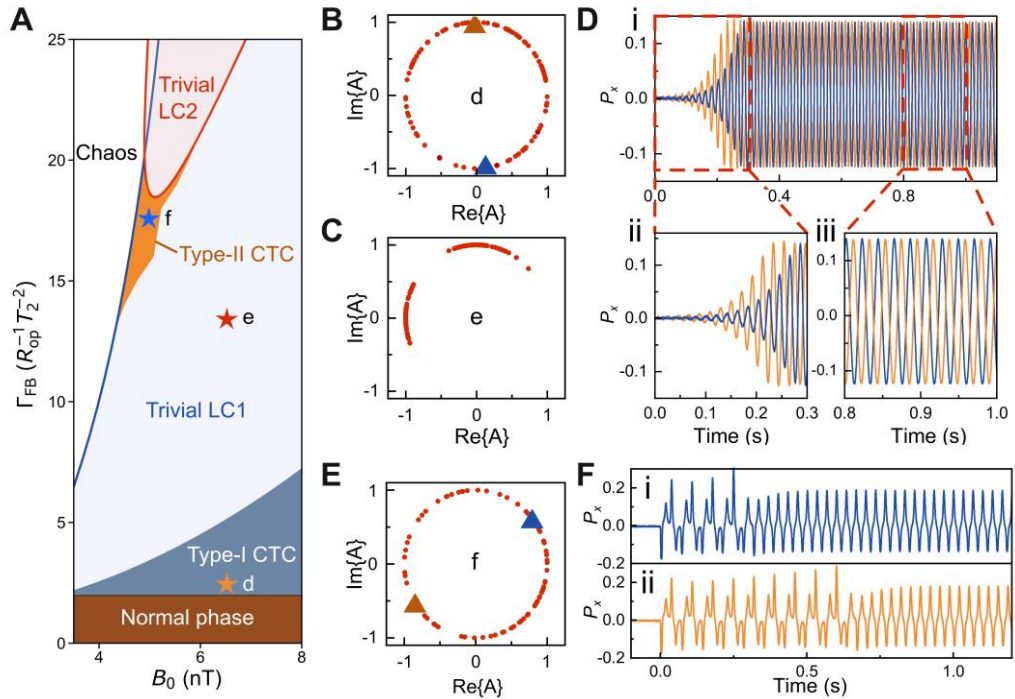

**Fig.3. Two types of time crystal phases.** (**A**) Phase diagram showing type-I (gray region) and type-II CTCs (orange region). (**B**) Normalized complex amplitude for the type-I CTC (point d in Fig.3A). The data is distributed over the unit circle, indicating a phase distribution of $2\pi$. (**C**) Normalized complex amplitude for the trivial LC1 phase (point e in Fig.3A). The data is distributed on two discontinuous arcs, which do not meet the $2\pi$ random phase required by the time crystal. **(D)** (i) Experimental output signal $P_x$ for two instances exhibiting a phase difference close to $\pi$ within the time crystal phase. They are represented by triangles in Fig. 3C. (ii) and (iii) Zoom in on the region 0-0.4 s and 0.8-1.2 s from Fig. 3D (i). **(E)** Normalized complex amplitude of the type-II CTC. The data has a 2pi phase distribution. **(F)** Two experimental results of the output signal $P_x$ induced by the type-II CTC. They are represented by triangles in Fig. 3E. After a short LC2 oscillation, the system remains in LC1 oscillation, and the different LC2 oscillation times experienced by (i) and (ii) result in a phase difference of $\pi$.

For our system, type-I CTCs can be found in the LC1 phase near the bifurcation from the normal phase to the limit cycle phase. As shown in Fig. 3A, the LC1 phase shows type-I CTC dynamics in the grey region near the normal phase, and shows trivial dynamics away from the bifurcation. The boundary between the trivial LC1 phase and the type-I CTC phase marks the critical point for the unstable manifold of unstable fixed point $\mathbf{P}^{(0)}$ to change dimensionality. When the feedback increases from the type-I CTC phase to the trivial LC1 phase, the real part of one eigenvalue of the linearized system at $\mathbf{P}^{(0)}$ changes from positive to negative, so that the dimension of the unstable manifold changes from 2 to 1. A detailed discussion is provided in Supplementary Materials. In the experiment, we set the magnetic field $B_0$ to $6.52\,\mathrm{nT}$ and activate the pump and probe lasers to attain the steady state $\mathbf{P}^{(0)}$, then turn on the feedback to activate the artificial spin-spin interactions. For comparative analysis, we adjust the feedback resistor to set the feedback factor $\Gamma_\mathrm{FB}$ at $2.4 R_\mathrm{op}^{-1} T_2^{-2}$ and $13.4 R_\mathrm{op}^{-1} T_2^{-2}$, corresponding to the time-crystal region (orange pentagram d) and non-time-crystal region (red pentagram e), respectively. Fig.3B and C show the normalized complex amplitudes of the spin polarization in the time-crystal and non-time-crystal phases. The complex amplitude in the time-crystal phase is uniformly distributed over the unit circle, indicating that the periodic motion has a random phase from $0$ to $2\pi$. In contrast, the complex amplitude in the non-time-crystal phase is distributed over two disconnected segments of the unit circle. Ideally, the phase should exhibit a two-point distribution, but system instability caused by temperature fluctuations of the gas cell broadens the phase distribution. We select two points (triangles) in Fig. 3B whose phase difference is close to $\pi$, and their $P_x$ output signals are shown in Fig.3D. Throughout $0.3\,\mathrm{s}$, the amplitude of both signals grows gradually, but the growth rates are different, resulting in a difference in the time required to achieve the same stable amplitude. After

$0.8\,\text{s}$, the phase difference between the two oscillations is close to $\pi$ as shown in Fig. 3D(ii).

Type-II CTCs occur near the pitchfork bifurcation of LC1 towards chaos, and close to LC2 phase, requiring that the orbit LC2 is unstable, but the instability is not too strong. Through numerical search, we can find the parametric region for type-II CTC phase as shown as the orange region in Fig. 3A. A detailed discussion is provided in Supplementary Material. The experiment uses a magnetic field of $B_0 = 4.98\,\text{nT}$ and a feedback gain of $\Gamma_{FB} = 17.6 R_{op}^{-1} T_2^{-2}$. A square wave signal with a frequency of $250\,\text{mHz}$ and a duty cycle of $85\%$ controls the feedback loop. The experiment runs 90 times, and Fig. 3F shows two examples of the results. The system starts from $\mathbf{P}^{(0)}$, temporarily experiences a phase of unstable LC2 oscillations with a bimodal structure before reaching a stable LC1 periodic orbit. The instability of LC2 leads to varying periods, which randomize the phase of the stable periodic motion. The phase is randomly distributed from $0$ to $2\pi$, as shown in Fig.3E, where the Fourier data of Fig. 3F are represented by two triangles.

Finally, we test the robustness of time crystals to persistent perturbations, which is one of the essential conditions for time crystals. We introduce quasi-Gaussian noise on the pump laser. The bandwidth of the quasi-Gaussian noise is set to 30 Hz near the resonance or 100 Hz and 1000 Hz away from the resonance. The noise intensity varies from $10\%$ to $50\%$. Figure 4 shows a comparison of the pump intensity without noise (blue line) and with a noise bandwidth of $30\,\text{Hz}$ and an intensity of $50\%$. Figures 4B and C display the results of type-I and type-II CTCs, respectively. The noise causes only a slight change in the amplitude of the time crystals. We analyze the variation of the time-crystals oscillation frequency under different bandwidths and noise intensities $\eta$. Figure 4D and E present the results of type-I and type-II CTCs, respectively. Both time crystals demonstrate strong robustness, with the frequencies changing slightly under noise. Significantly, the type-I CTCs are almost immune to noise.

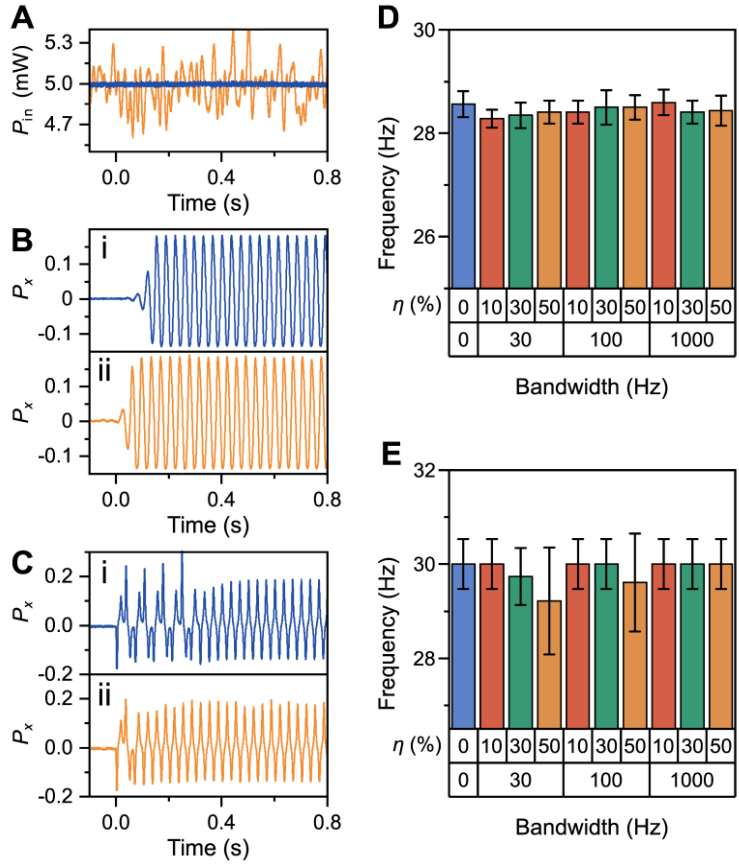

**Fig.4. Robustness to noise perturbations of CTCs. (A)-(C)** A comparison between the signals of the pump laser without noise (blue solid line) and with noise at the bandwidth of 30 Hz and intensity of 10% (orange solid line), where (A) pump laser power (B) type-I CTC signals (C) type-II CTC signals. **(D)-(E)** Oscillation frequencies of the type-I and type-II CTCs signals at different noise bandwidths and intensities $\eta$, respectively.

In summary, we have revealed the spontaneous symmetry breaking in continuous time crystals. Two different mechanisms are proposed to realize CTCs with random phases, which are protected by manifold topology and near-chaotic hopping, respectively. In the type-I CTC, the connected topology of the unstable manifold of the unstable fixed points plays an important role, and in the type-II CTC, the near-chaotic hopping between unstable periodic orbits, which amplifies the initial fluctuation, is the main origin of the random phase. We have synthesized spin-spin nonlinear interactions in thermal atomic ensembles based on a measurement-feedback method, experimentally observed the nonlinear dynamics including limit cycles and chaos, and observed both type-I and type-II CTCs, showing extraordinary robustness to noises. Our work provides general recipes for the realization of CTCs, and can stimulate the development of novel techniques, such as long-lived oscillation generation and robust precision metrology.

**References and Notes**

1. A. Shapere, F. Wilczek, *Phys. Rev. Lett.* **109**, 160402 (2012).

2. F. Wilczek, *Phys. Rev. Lett.* **109**, 160401 (2012).

3. T. Li *et al.*, *Phys. Rev. Lett.* **109**, 163001 (2012).

4. F. Wilczek, *Phys. Rev. Lett.* **111**, 250402 (2013).

5. P. Bruno, *Phys. Rev. Lett.* **111**, 070402 (2013).

6. H. Watanabe, M. Oshikawa, *Phys. Rev. Lett.* **114**, 251603 (2015).

7. V. K. Kozin, O. Kyriienko, *Phys. Rev. Lett.* **123**, 210602 (2019).

8. Z. Gong, R. Hamazaki, M. Ueda, *Phys. Rev. Lett.* **120**, 040404 (2018).

9. F. Iemini *et al.*, *Phys. Rev. Lett.* **121**, 035301 (2018).

10. F. M. Gambetta, F. Carollo, M. Marcuzzi, J. P. Garrahan, I. Lesanovsky, *Phys. Rev. Lett.* **122**, 015701 (2019).

11. B. Buča, J. Tindall, D. Jaksch, *Nat. Commun.* **10**, 1730 (2019).

12. G. Buonaiuto, F. Carollo, B. Olmos, I. Lesanovsky, *Phys. Rev. Lett.* **127**, 133601 (2021).

13. M. Hajdušek, P. Solanki, R. Fazio, S. Vinjanampathy, *Phys. Rev. Lett.* **128**, 080603 (2022).

14. M. Krishna, P. Solanki, M. Hajdušek, S. Vinjanampathy, *Phys. Rev. Lett.* **130**, 150401 (2023).

15. M. P. Zaletel *et al.*, *Rev. Mod. Phys.* **95**, 031001 (2023).

16. S. Choi *et al.*, *Nature*. **543**, 221–225 (2017).

17. S. Autti, V. B. Eltsov, G. E. Volovik, *Phys. Rev. Lett.* **120**, 215301 (2018).

18. J. Rovny, R. L. Blum, S. E. Barrett, *Phys. Rev. Lett.* **120**, 180603 (2018).

19. J. Smits, L. Liao, H. T. C. Stoof, P. van der Straten, *Phys. Rev. Lett.* **121**, 185301 (2018).

20. M. Ippoliti, K. Kechedzhi, R. Moessner, S. L. Sondhi, V. Khemani, *PRX QUANTUM*. **2**, 030346 (2021).

21. H. Keßler *et al.*, *Phys. Rev. Lett.* **127**, 043602 (2021).

22. A. Kyprianidis *et al.*, *Science*. **372**, 1192-+ (2021).

23. J. Randall *et al.*, *Science*. **374**, 1474–1478 (2021).

24. A. Scheie *et al.*, *Nat. Commun.* **13**, 5796 (2022).

25. X. Zhang *et al.*, *Nature*. **607**, 468–473 (2022).

26. W. Beatrez *et al.*, *Nat. Phys.* **19**, 407–413 (2023).

27. H. Taheri, A. B. Matsko, L. Maleki, K. Sacha, *Nat. Commun.* **13**, 848 (2022).

28. B. Wang *et al.*, *Laser Photonics Rev.* **16**, 2100469 (2022).

29. W. Zhu, H. Xue, J. Gong, Y. Chong, B. Zhang, *Nat. Commun.* **13**, 11–11 (2022).

30. J. Zhang *et al.*, *Nature*. **543**, 217–220 (2017).

31. P. Frey, S. Rachel, *Sci. Adv.* **8**, eabm7652 (2022).

32. X. Mi *et al.*, *Nature*. **601**, 531–536 (2022).

33. N. Y. Yao, C. Nayak, L. Balents, M. P. Zaletel, *Nat. Phys.* **16**, 438–447 (2020).

34. N. Y. Yao, A. C. Potter, I.-D. Potirniche, A. Vishwanath, *Phys. Rev. Lett.* **118**, 030401 (2017).

35. R. Moessner, S. L. Sondhi, *Nat. Phys.* **13**, 424–428 (2017).

36. D. V. Else, B. Bauer, C. Nayak, *Phys. Rev. X.* **7**, 011026 (2017).

37. V. Khemani, A. Lazarides, R. Moessner, S. L. Sondhi, *Phys. Rev. Lett.* **116**, 250401 (2016).

38. S. Autti *et al.*, *Nat. Commun.* **13**, 3090 (2022).

39. P. Kongkhambut *et al.*, *Science*. **377**, 670–673 (2022).

40. T. Liu, J.-Y. Ou, K. F. MacDonald, N. I. Zheludev, *Nat. Phys.* **19**, 986–991 (2023).

41. X. Wu *et al.*, *Nat. Phys.* (2024), doi:10.1038/s41567-024-02542-9.

42. I. Carraro-Haddad *et al.*, *Science*. **384**, 995–1000 (2024).

43. A. Greilich *et al.*, *Nat. Phys.* **20**, 631–636 (2024).

44. Y.-H. Chen, X. Zhang, *Nat. Commun.* **14**, 6161 (2023).

45. H. Keßler, J. G. Cosme, M. Hemmerling, L. Mathey, A. Hemmerich, *Phys. Rev. A*. **99**, 053605 (2019).

46. M. Jiang, H. Su, Z. Wu, X. Peng, D. Budker, *Sci. Adv.* **7**, eabe0719 (2021).

47. Y. Tang *et al.*, *Phys. Rev. Lett.* **130**, 193602 (2023).

**Acknowledgments:**

We thank Prof. Li You for helpful discussions.

**Funding:** This work is supported by the National Key R&D Program of China (2023YFA1407600), the National Natural Science Foundation of China (NSFC) (12275145, 92050110, 12274031, 62005145, 91736106, 11674390, and 91836302), New Cornerstone Science Foundation, and the Research Grants Council of Hong Kong (AoE/P-502/20 and 17309021).

**Author contributions:** Y.-C.L, Y.-J.T. and C.-Y.W. conceived the project. C.-Y.W. and Y.-J.T. performed theoretical analysis. Y.-J.T., B.L. and J.P. performed the experiments and all the authors analyzed the data. Y.-J.T., C.-Y.W., Y.-C.L and S. Z wrote the paper with input from all authors. All authors contributed to the discussion and review the manuscript. Y.-C.L and S.Z. supervised the projects.